\documentclass[letterpaper,twocolumn,10pt]{article}
\usepackage{usenix-2020-09}
\pdfoutput=1

\usepackage{findlay}
\microtypecontext{spacing=nonfrench}

\def\bpfcontain{\textsc{BPFContain}}
\date{}

\title{\Large \bpfcontain: Fixing the Soft Underbelly of Container Security}

\author{
{\rm William Findlay}\\
Carleton University
\and
{\rm David Barrera}\\
Carleton University
\and
{\rm Anil Somayaji}\\
Carleton University
}

\begin{document}

\maketitle

\begin{abstract}
Linux containers currently provide limited isolation guarantees.  While containers separate namespaces and partition resources, the patchwork of mechanisms used to ensure separation cannot guarantee consistent security semantics.  Even worse, attempts to ensure complete coverage results in a mishmash of policies that are difficult to understand or audit.  Here we present \bpfcontain, a new container confinement mechanism designed to integrate with existing container management systems.  \bpfcontain{} combines a simple yet flexible policy language with an eBPF-based implementation that allows for deployment on virtually any Linux system running a recent kernel.  In this paper, we present \bpfcontain's policy language, describe its current implementation as integrated into docker, and present benchmarks comparing it with current container confinement technologies.
\end{abstract}

\section{Introduction}

Linux containers have become the preferred unit of application management in the cloud, forming the foundation of Docker~\cite{docker}, Kubernetes~\cite{kubernetes}, Snap~\cite{snap}, Flatpak~\cite{flatpak}, and others.  By including just the binaries, libraries, and configuration files needed by an application, containers enable simplified deployment of vendor-packaged applications, rapid horizontal application scaling, and direct developer-to-production DevOps workflows, all without the overhead of hypervisor-based virtual machines (HVMs).

Many have recognized that containers currently offer weak isolation guarantees ~\cite{sultan2019_container_security, sun2018_security_namespace, xin2018_container_security}.  Weak container confinement is less of a risk when all containers on a given host are deployed by the same party.  However, strong container isolation mitigates privilege escalation attacks and is a critical requirement in multi-tenancy environments where attacker-controlled containers co-exist with benign containers.

The key to improving container isolation lies in recognizing that isolation is not the same as virtualization.  Virtualization can exhibit isolation characteristics as a side effect; however, such isolation is rarely enough on its own to serve a security barrier.  In networking, network address translation (NAT) virtualizes IP addresses so one IP address can be shared by an entire network of systems.  While this many-to-one mapping provides a significant degree of isolation to systems behind the NAT, it is no substitute for a network firewall, especially one that is configured, e.g., to block outbound connections.  Linux containers are virtualized using cgroups~\cite{cgroups} and namespaces~\cite{namespaces}, while confinement is enforced through seccomp-bpf~\cite{seccomp_bpf} and mandatory access control mechanisms such as SELinux~\cite{smalley2001_selinux} and AppArmor~\cite{cowan2000_apparmor}.  While these confinement mechanisms are powerful and flexible, container isolation was not their primary design goal, and currently they can only accomplish the task with complex policies that are difficult to write and audit.

To properly isolate containers, the kernel requires clear rules about
what interactions are permissible, whether between containers or with
the host OS.  Much like firewall rules specify which packets may
traverse it, OSes need unambiguous, auditable rules to determine what
kernel-level operations are and are not allowed based on container
boundaries.  Following Unix tradition, the Linux kernel provides only security mechanisms and abstractions for defining security policies, but leaves policies themselves to be managed in userspace. However, unlike processes, files, and network connections, the Linux kernel has no unified abstraction around containers.

We assert that the lack of strong isolation guarantees for containers arises from a semantic gap between the security mechanisms that currently exist in the Linux kernel and the policies that we wish to define to isolate containers.  As long as the kernel does not implement container-level access control mechanisms, the result will be complex, circumventable container isolation.

The Linux kernel has recently gained an alternative way to implement security abstractions: eBPF.  An extension of the Berkeley Packet Filter~\cite{mccanne1993_bpf}, Linux's eBPF now allows for complex monitoring and manipulation of kernel-level events. Thanks to implicit load-time verification of eBPF bytecode, eBPF provides strong performance, portability, and safety guarantees.  More recent improvements to eBPF~\cite{corbet2019_krsi} have enabled interfacing with Linux Security Module (LSM) hooks, allowing eBPF code to implement new kernel-level security mechanisms.

This paper proposes \bpfcontain{}, a novel approach to container security under the Linux kernel, rectifying overprivileged and insecure containers. Leveraging eBPF, \bpfcontain{} uses runtime security instrumentation to implement container-aware policy enforcement and harden the host kernel against privilege escalation attacks mounted from within containers. Specifically, \bpfcontain{} attaches eBPF programs to LSM hooks and critical functions within the kernel to enforce per-container policy in kernelspace.

Thanks to eBPF's dynamic instrumentation capabilities, this integration occurs at runtime and requires no modification or patching of the kernel.  \bpfcontain{} addresses the container userspace/kernelspace semantic gap by defining a new YAML-based policy language for container confinement that allows for simple default deny and default allow policies, high-level semantically meaningful permissions, and (where necessary) fine-grained control over the LSM API, all at the level of containers.

While \bpfcontain{} can co-exist with seccomp-bpf, SELinux, and other existing Linux security mechanisms, it does not rely on them, and in fact in the context of container isolation, \bpfcontain{} makes them redundant.  It also has modest userspace requirements, consisting only of a small daemon and a control program, both of which are written in Rust. In summary, we make the following contributions:

\begin{itemize}
  \item We present the design, implementation, and evaluation of \bpfcontain{}, a container-aware security enforcement mechanism for Linux. \bpfcontain{} is available\footnote{\url{https://github.com/willfindlay/bpfcontain-rs}} under a GPLv2 license, and installation requires only a 5.10 or newer Linux kernel.  While there is past work on using eBPF for process sandboxing~\cite{findlay2020_bpfbox}, \bpfcontain{} is both more general and more deployable, implementing mechanisms and a policy language specific to container confinement and built using tools such as BPF CO-RE and Rust that have much lower space and runtime overhead.

\item We design a flexible policy language for confining containers that offers optional layers of granularity to meet a wide range of real world container use cases. The policy language is YAML-based, and expressive enough that it can be used to confine individual system resources, yet simple enough that it affords ad-hoc confinement use cases.

  \item We discuss integrating \bpfcontain{} with existing container management frameworks without modifying their source code, offering significant security advantages over traditional approaches to container isolation and least-privilege.
\end{itemize}

The rest of this paper proceeds as follows. \Cref{sec:background} presents background and our motivation for implementing an eBPF-based container security mechanism. \Cref{sec:bpfcontain} presents an overview of \bpfcontain{} and discusses our threat model and design goals. \Cref{sec:policy} describes \bpfcontain{}'s policy language. \Cref{sub:implementation} discusses the design and implementation of its userspace components and enforcement engine. \Cref{sec:eval} presents an evaluation of \bpfcontain{}'s security and performance. \Cref{sec:related} covers related work and \Cref{sec:discussion} discusses limitations and opportunities for future work. \Cref{sec:conclusion} concludes.

\section{Background and Motivation}\label{sec:background}

This section reviews current techniques for achieving virtualization, isolation, and least privilege on Linux containers. It also provides background on the classic and extended Berkeley Packet Filters (BPF and eBPF), and discusses the motivation behind a new container-focused security enforcement mechanism.

\begin{figure}[t]
\includegraphics[width=0.8\linewidth]{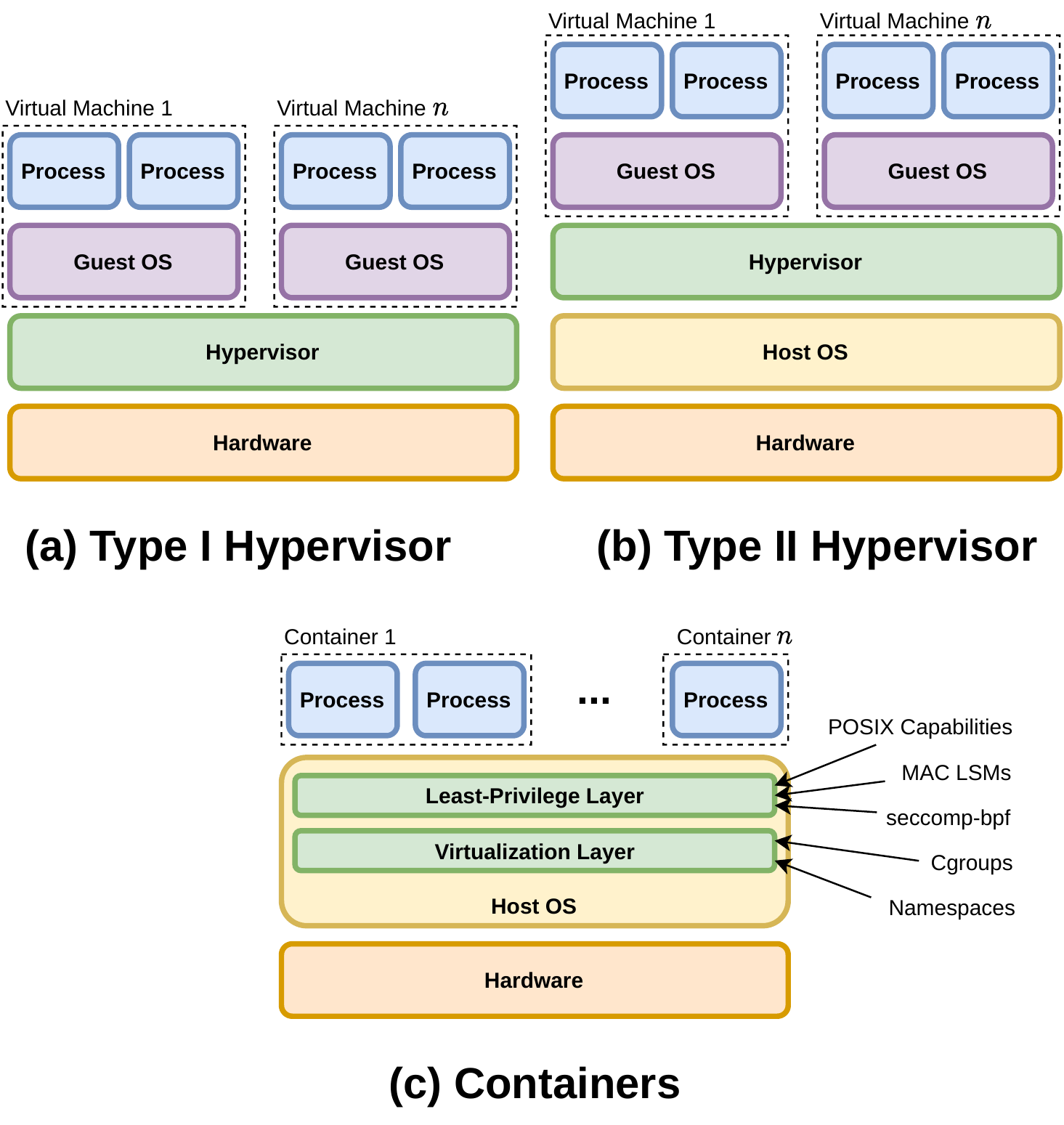}
  \caption{
    Comparison of virtual machine and container architectures. Type I hypervisors \textbf{(a)} virtualize and control the underlying hardware directly, but require full guest operating systems on top of the virtualization layer. Type II hypervisors \textbf{(b)} run on top of a host operating system but still require full guest operating systems above the virtualization layer. Containers \textbf{(c)} achieve virtualization using a thin layer provided by the host operating system itself. They share the underlying operating system kernel and resources, requiring no guest operating system.
  }\label{fig:virt}
\end{figure}

A container is a userspace representation of a set of processes that share the same virtualized view of files and system resources. Containers run directly on the host operating system and share the underlying OS kernel, and thus do not require a full guest operating system or hypervisor to provide the virtualized view of the system to applications (see \Cref{fig:virt}). To support security beyond basic process isolation, modern container runtimes rely on a variety of low-level Linux kernel facilities to enforce virtualization, isolation, and least-privilege. Note that by relying on a disparate set of mechanisms and corresponding policies, a failure in the enforcement or policy in any single mechanism exposes the container to attacks.

\subsection{Container Security}
\label{subsection:containers}

\paragraph*{Namespaces and Control Groups (cgroups).} These mechanisms allow for further confinement of processes by restricting the system resources that a process or group of processes can access. Namespaces isolate access by providing a process group a private, virtualized naming of a class of resources, such as process IDs, filesystem mountpoints, and user IDs. Cgroups (control groups) limit available \textit{quantities} of system resources, such as CPU, memory, and block device I/O. Namespaces and cgroups on their own cannot enforce fine-grained access policies (e.g., to resources within a namespace), and thus require additional mechanisms as described below.

\paragraph*{POSIX Capabilities.} These provide a finer-grained alternative to the all-or-nothing superuser privileges required by certain applications\cite{posix_capabilities,corbet2006_capabities_a,corbet2006_capabities_b}. POSIX capabilities can be used to grant limited additional privileges to specific processes, overriding existing discretionary permissions. Further, a privileged process may \textit{drop} specific capabilities that it no longer needs, retaining those it does. POSIX capabilities add complexity to the already complex Linux permission model \cite{corbet2006_capabities_b,corbet2006_capabities_a}. It is challenging to create a default set of capabilities that work for most use cases. For example, Docker provides containers with 15 Linux capabilities by default, including \texttt{CAP\_DAC\_OVERRIDE}, which allows a container to override all discretionary access control checks \cite{sultan2019_container_security,docker}.

\paragraph*{System Call Filtering.} Linux's seccomp-bpf \cite{seccomp_bpf} is a common approach for reducing the set of system calls available to an application. Container runtimes will often ship with a seccomp-bpf policy that prevents containers from making calls that are known to be dangerous. Despite the high degree of control that seccomp-bpf offers to applications, it is not without its own usability and security concerns. seccomp-bpf policy is easy to misconfigure, resulting in potential security violations; for instance, an attacker may entirely circumvent a policy that specifies restrictions on the \texttt{open(2)} system call but not \texttt{openat(2)}.

\paragraph*{Mandatory Access Control (MAC).} The Linux Security Modules (LSM) API~\cite{wright2002_lsm} provides an extensible security framework for the Linux kernel, allowing for the implementation of kernelspace security mechanisms that can be chained together. SELinux~\cite{smalley2001_selinux} and AppArmor~\cite{cowan2000_apparmor} use the LSM API to provide fine-grained mandatory access control throughout the system. Container runtimes often ship default MAC policies for their containers, and requesting enforcement if such a MAC system is enabled. To our knowledge, no container runtime will refuse to run if no LSM is loaded, effectively failing open. MAC policies are also known to be difficult to maintain and audit~\cite{schreuders12_towards}.

\subsection{Classic and Extended BPF}

The original Berkeley Packet Filter (BPF) \cite{mccanne1993_bpf}, hereafter referred to as Classic BPF, was a packet filtering mechanism implemented initially for BSD Unix.
Classic BPF implemented a simple register-based virtual machine language and efficient buffer data structures to minimize the context switches while making filtering decisions. As an efficient packet filtering mechanism, Classic BPF quickly gained traction in the Unix community and has since been ported to many Unix-like operating systems, most notably Linux~\cite{linux_bpf}, OpenBSD~\cite{openbsd_bpf}, and FreeBSD~\cite{freebsd_bpf}. On Linux, the seccomp sandboxing facility has been extended to use BPF to make security decisions about system calls in a confined process (c.f., seccomp-bpf~\cite{drewry2012_seccomp_bpf,seccomp_bpf}).

A complete rewrite of the Linux BPF engine, dubbed Extended BPF (eBPF), was merged into the mainline kernel~\cite{starovoitov2014_ebpf} in 2014. eBPF expands on the original BPF specification by introducing: an extended instruction set, 11 registers (10 of which are general-purpose), access to allow-listed kernel helpers, Just-in-time (JIT) compilation to native instruction sets, a program safety verifier, specialized data structures, and new program types which can be attached to a variety of system events in both userspace and kernelspace.

These extensions to the Classic BPF engine effectively turn eBPF into a general-purpose execution engine in the kernel with system introspection and kernel extension capabilities. eBPF programs execute in the kernel but are limited by a restricted execution context and pre-checked for safety by an in-kernel verification engine. In particular, eBPF programs are limited to a 512-byte stack, cannot access unbounded memory regions, must not have back-edges in their control flow, and must provably terminate~\cite{gregg2019_bpf}. As a consequence of these restrictions, eBPF programs are not Turing-complete. Where necessary, an eBPF program can make calls to a set of allow-listed kernel helpers to obtain additional functionality, such as access to external memory regions and various kernel facilities such as signalling or random number generation~\cite{gregg2019_bpf}.

\begin{figure}[tb]
  \centering
  \includegraphics[width=0.8\linewidth]{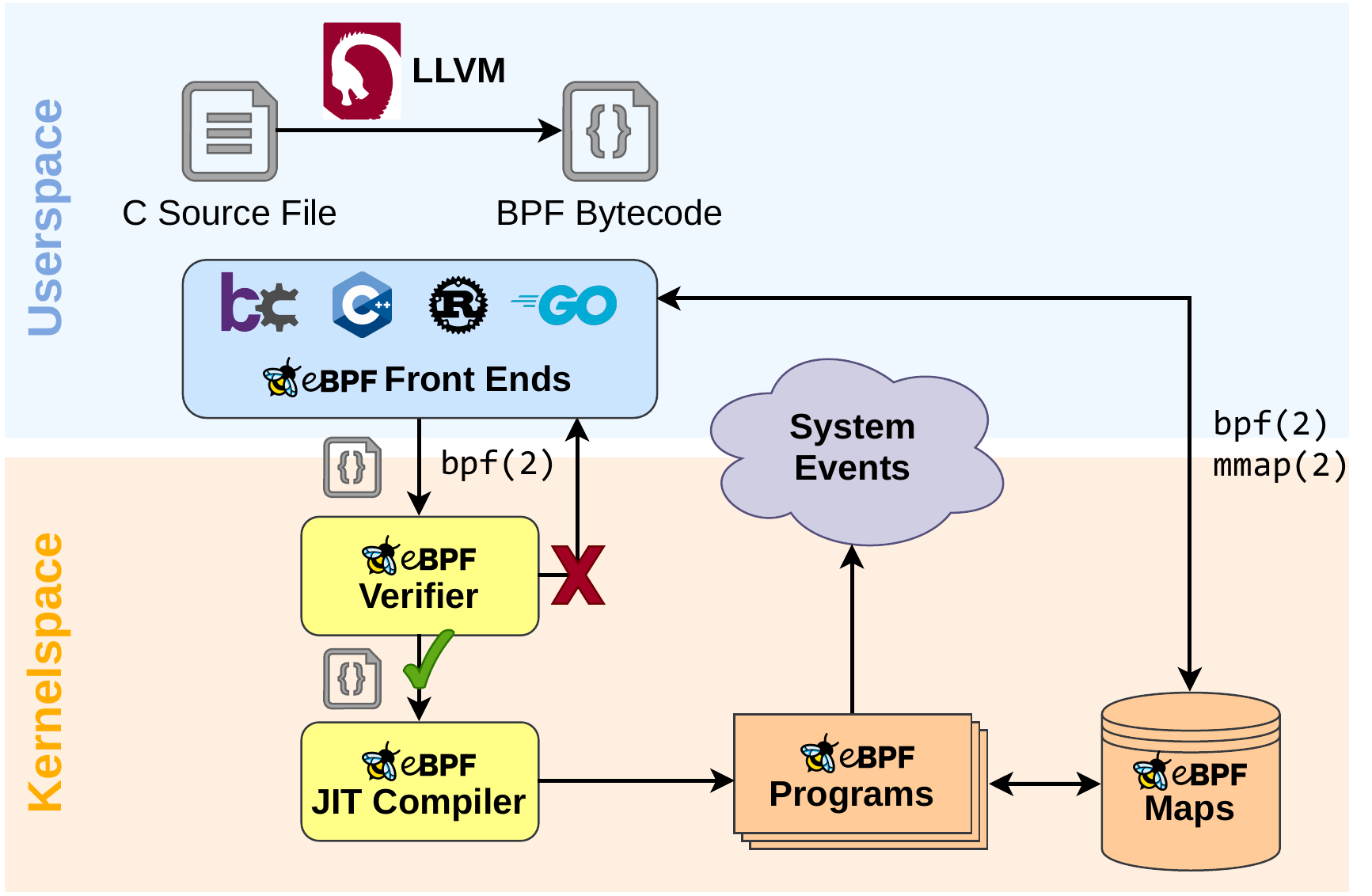}
  \caption{
    Linux kernel eBPF architecture.
}\label{fig:ebpf}
\end{figure}

A privileged userspace process may load an eBPF program into the kernel using Linux's \texttt{bpf(2)} system call (see \Cref{fig:ebpf}). While it is possible to write eBPF bytecode by hand \cite{gregg2019_bpf}, several front-ends exist for compiling eBPF bytecode from a restricted subset of the C programming language\footnote{In principle, this language need not be C. For instance, a framework exists for writing eBPF programs in pure Rust \cite{redbpf}. However, C is a popular choice since it is tightly coupled with the underlying implementation of the kernel.}, including bcc \cite{bcc} and libbpf \cite{libbpf}. These front-ends typically use the LLVM \cite{llvm_bpf} compiler toolchain to produce BPF bytecode. When the kernel receives a request to load an eBPF program, it first checks the bytecode to ensure that it conforms to the safety invariants outlined above. If the verifier accepts the program, it may then be attached to one or more system events. When an event fires, the eBPF program is executed via just-in-time compilation to the native instruction set. eBPF programs can store data in several specialized in-kernel data structures, made accessible to userspace via the \texttt{bpf(2)} system call or a direct memory mapping.

\subsection{Motivation}\label{sec:motivation}

Isolation has not been a focus of container management frameworks, with many taking a lax attitude towards least-privilege enforcement~\cite{sultan2019_container_security}. Popular container frameworks such as Docker~\cite{docker} provision overly-permissive default access, rely on a complex and often ill-suited suite of security mechanisms provided by the host system (see~\Cref{subsection:containers}), and support insecure configuration options~\cite{docker,sultan2019_container_security,xin2018_container_security,findlay2020_bpfbox}. Again, the main challenge in providing strong container isolation is that the kernel has no unified abstraction to represent containers, which results in a patchwork of security mechanisms (see~\Cref{subsection:containers}) each enforcing specific policies. A vulnerability in any individual mechanism, or a misconfiguration in any individual policy opens up the container or system to attack.

We argue that the key to providing strong container isolation is to bridge the semantic gap between the kernel security enforcement mechanisms and container-level security policies. One approach is to extend the Linux kernel with a container-aware LSM, much like AppArmor or SELinux (see \Cref{subsection:containers}). However, maintaining out-of-tree kernel modules is challenging, as they must be continually updated as the kernel evolves. Another issue is adoption; users may be reluctant to run a third-party module because bugs can cause system crashes or data loss.

In the case of container isolation the situation is even worse, as a single kernel-level security mechanism is unlikely to work for all use cases.  Some users (e.g., operators of multi tenant container clouds) will want strong isolation for their containers, while others (e.g., end users) will want containers to deeply integrate with the resources provided by other containers and the host system.

Containers have strong conceptual backing in userspace, but lack a unifying abstraction in the kernel. From the kernel's perspective, a containerized process is just like any other---it may be running under a different set of namespaces or in a different control group, but there exists no unifying definition of precisely what a container is, from the kernel's perspective. This lack of a solid container abstraction in turn widens the semantic gap between per-container security policy authored in userspace and policy enforcement in the kernel. We argue that \bpfcontain{} can provide such a unifying abstraction for policy enforcement. Since \bpfcontain{} requires no modification to the host kernel and can be dynamically loaded at runtime, we can provide such an abstraction without necessarily sacrificing forward compatibility with other approaches.

In summary, \bpfcontain{} came out of the realization that 1) with eBPF, we had the technology to implement specialized security abstractions and that 2) container confinement was a problem that could use a problem-specific security abstraction, given the complexity of existing solutions.

\section{\bpfcontain{}}\label{sec:bpfcontain}

\begin{figure}[h]
  \centering
  \includegraphics[width=.95\linewidth]{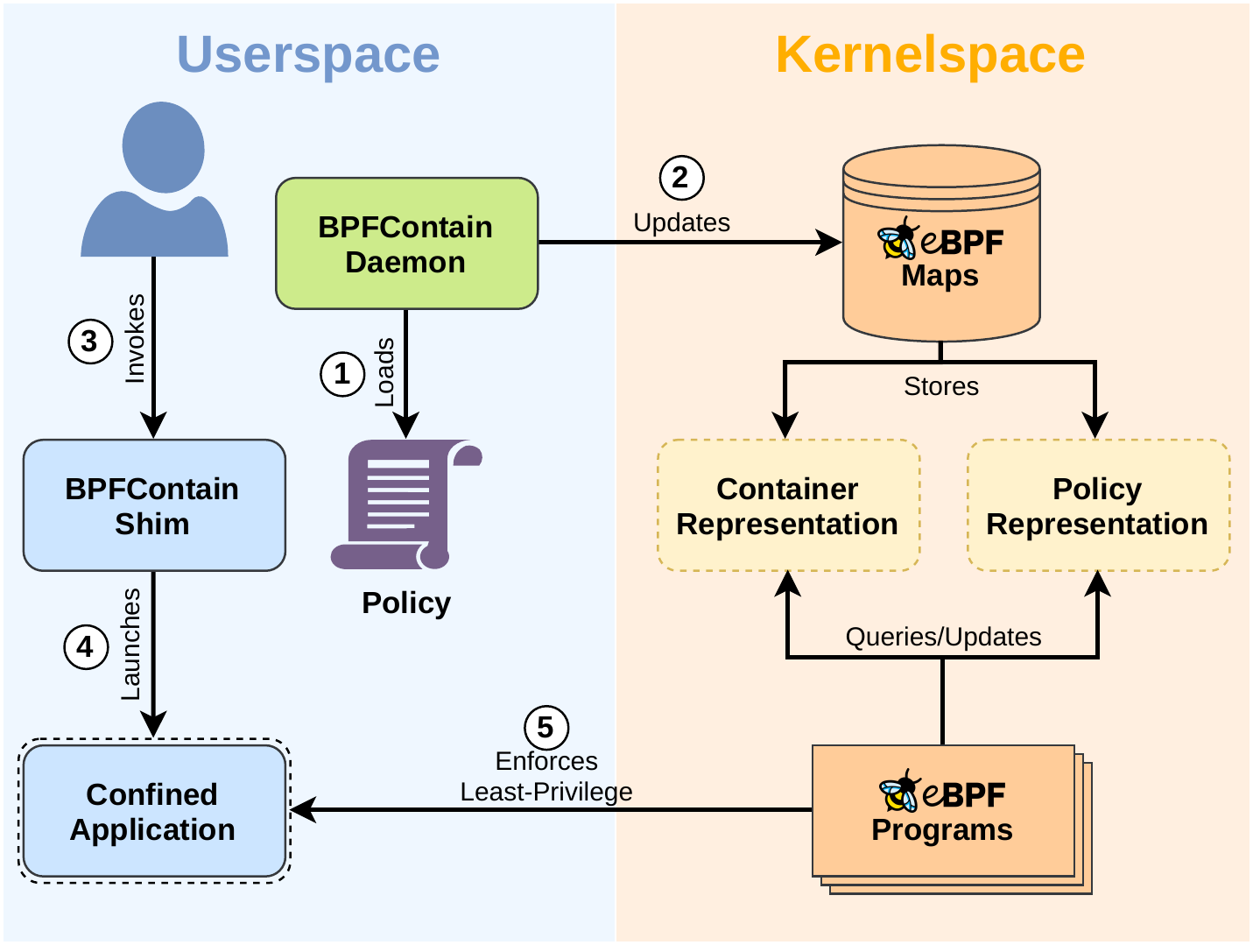}
  \caption{A high-level depiction of how users can use \bpfcontain{} to enforce least privilege on an application. In userspace, a privileged daemon loads parses policy files and loads policy into the kernel. The user invokes a wrapper command to launch the confined application. In the kernel, eBPF maps track the state of confined applications and active policy. eBPF LSM programs then enforce this policy when a confined application requests access to a mediated system resource.}\label{fig:confining}
\end{figure}

To confine applications, \bpfcontain{} leverages eBPF~\cite{starovoitov2014_ebpf}, a Linux technology that allows privileged userspace processes to make verifiably safe runtime extensions to the operating system kernel. A privileged userspace daemon manages \bpfcontain{}'s eBPF components and loads policy into the kernel. When a user wishes to confine an application, they invoke a shim wrapper command, which associates itself with a specific \bpfcontain{} policy and subsequently executes the target application.  \Cref{fig:confining} illustrates this process at a high level.

\bpfcontain{} associates a confined application and all its children with a universally unique ID, called the \enquote{container ID}, which is used to track various information, including namespace membership and policy association. eBPF maps store per-container information and security policies, which are queried at runtime by eBPF programs responsible for making policy decisions. In this way, \bpfcontain{} makes the kernel \textit{container-aware} and provides a mechanism for enforcing strong least-privilege guarantees at the granularity of individual containers.

\subsection{Threat Model}
\label{sub:threat-model}

\bpfcontain{} aims to protect against two categories of container-related attacks: (1) inter-container attacks, where one container attempts to interfere with or take over another; and (2) containers attacking the host, either by escaping confinement or launching denial of service or resource consumption attacks. Intra-container attacks, where an application inside a container might attack the container itself are not our focus, and are better addressed by deploying existing security mechanisms (see~\Cref{subsection:containers}) inside the container. Host-to-container attacks are also out of scope, as they likely require the use of hardware security mechanisms~\cite{sultan2019_container_security,xin2018_container_security,chen2019_container_dos}.

We consider an attacker who can create and deploy containers to a multi-tenant shared container cloud. While the attacker has full control over the contents of each container they create, the attacker does not have administrative privileges on the host. The attacker's goal is to attack co-located containers or the host system.  Such attacks can have a variety of goals including information extraction, application compromise, or denial of service.

We assume that an attacker-controlled process must make use of kernel interfaces (i.e., system calls) in order to mount an attack against other containers or the host system.  We also assume that appropriate resource bounds are ensured through proper cgroups configuration in order to prevent resource-based denial of service attacks.  Also, attacks targeting the hardware (such as Spectre~\cite{kocher2019_spectre} and Rowhammer~\cite{mutlu2019rowhammer}) are outside the scope of \bpfcontain{}.

While this threat model might appear limited, it is the same one assumed by standard Linux kernel security mechanisms including SELinux and AppArmor.

\subsection{Design Goals}

We followed five design goals when designing both \bpfcontain{}'s policy language and
enforcement engine. These goals are enumerated below. Note that our focus is on isolation rather than virtualization of system resources, and \bpfcontain{} uses existing kernel facilities for virtualization. We discuss potential avenues for providing standalone virtualization capabilities in \bpfcontain{} in \Cref{subsection:ociintegration}.

\paragraph*{Security.}
\bpfcontain{} should be built from the ground up with security in mind. In particular, security should not be an opt-in feature and \bpfcontain{} should adhere to the principle of least privilege~\cite{saltzer1975_protection} by default. It should be easy to tune a \bpfcontain{} policy to respond to new threats or configurations.

\paragraph*{Simplicity.}
The \bpfcontain{} policy language should be simple enough for ad-hoc use cases without sacrificing security. In pursuit of this goal, our policy language takes inspiration from other high-level policy languages for containerized applications, such as Snapcraft~\cite{snap}. We also base our policy language syntax on the YAML specification~\cite{yaml}, which is both well-understood and widely-used as a configuration language.

\paragraph*{Flexibility.}
When studying typical use cases for container technology, it quickly became apparent that there exists a dichotomy in the way containers are used in practice. In industry, containers tend to be used to deploy composable micro-services, particularly in cloud contexts~\cite{amaral2015performance}. On the other hand, they are often used to isolate desktop applications~\cite{snap, flatpak}. In light of these findings, we designed \bpfcontain{} to work in \textit{both} use cases, providing a means of securing micro-services and restricting desktop applications' behaviour.

\paragraph*{Transparency.}
Confining an application using \bpfcontain{} should not require modifying the application's source code or running the application using a privileged SUID (Set User ID root) binary. \bpfcontain{} should be entirely agnostic to the rest of the system and should not interfere with its regular use.

\paragraph*{Adoptability.}
\bpfcontain{} should be adoptable across various system configurations and should not negatively impact the running system. It should be possible to deploy \bpfcontain{} in a production environment without impacting system stability and robustness or exposing the system to new security vulnerabilities. \bpfcontain{} relies on the underlying properties of its eBPF implementation to achieve its adoptability guarantees.

\section{\bpfcontain{} Policy}\label{sec:policy}

\bpfcontain{} exposes a YAML-based policy configuration language to system administrators. By default, the \bpfcontain{} daemon loads policy from \texttt{/var/lib/bpfcontain/policy}, although this setting can be changed via an environment variable when running the daemon. At a minimum, a policy file consists of a unique \textit{name} (later used to compute a unique identifier for the policy) and a default \textit{entrypoint}, i.e.~a pathname to an executable along with optional arguments. For instance, consider a minimal policy file for a simple, statically linked program that reads from standard in and writes to standard out. This policy is depicted in \Cref{lst:minimal_policy}.

\begin{listing}[language=yaml,label=lst:minimal_policy,caption={A minimal \bpfcontain{} policy file for a statically linked application that reads from stdin and writes to stdout.}]
name: hello_minimal
entry: /usr/bin/hello.static
allow:
  # Grant read and write access to
  # /dev/tty* and /dev/pts/* devices
  - tty: rw
\end{listing}

In accordance with the principle of least privilege~\cite{saltzer1975_protection}, \bpfcontain{} policy defaults to a default-deny. This means that, with no additional information, the policy depicted in \Cref{lst:minimal_policy} would deny access to \textit{all} security-sensitive resources on the system, including regular files, directories, kernel interfaces such as character devices and special filesystems, network communication, interprocess communication, and POSIX capabilities. In other words, any process tied to this policy file would only be able to perform basic computational tasks, with no access to any resources gated by the operating system's reference monitor.

However, it is possible to specify that a default-allow policy be enforced instead. In this way, end-users can write a simple policy restricting \textit{specific} application behaviour without worrying about the details of writing a rigorous security policy, should they so choose. For instance, a user might wish to restrict an application's access to a specific subset of kernel interfaces without worrying about other access control decisions such as access to shared libraries. Since default-allow enforcement is strictly an opt-in feature, we can support such flexible confinement without exposing an unsuspecting user to any additional risk. Configuring our above example to be default allow is as simple as adding a new line with \texttt{default: allow}.

\bpfcontain{} supports three main categories of policy rules. \textit{Allow rules} grant access to system resources, \textit{deny rules} restrict access to system resources, and \textit{taint rules} specify conditions under which a \bpfcontain{} container should become \textit{tainted}. When a security policy specifies taint rules, the resulting container is considered \textit{untainted} until the taint rule is matched. Untainted containers are exempted from default-deny enforcement, meaning that it is possible to define a security policy that behaves \textit{as if it were default allow} during some predetermined setup phase. A policy without any taint rules is assumed to be tainted by default. Once tainted, it is impossible for a process to become untainted.

The concept of tainting both greatly simplifies and hardens the resulting security policy. This stems from the observation that the initial setup phase and main work loop of a given application often have totally disparate access patterns---for instance, a dynamically linked C application will map shared libraries into executable memory during its setup phase, but is unlikely to perform similar mappings for the remainder of its lifecycle. Eliminating these initial access patterns from security policy simplifies policy authorship while simultaneously preventing such access patterns once a process enters its main work loop. Using this notion of tainting, we can consider a revised policy for a dynamically compiled version of our simple application, depicted in \Cref{lst:taint_policy}.

\begin{listing}[language=yaml,label=lst:taint_policy,caption={A \bpfcontain{} policy file for a dynamically linked application that reads from stdin and writes to stdout. Policy enforcement begins after the first read from stdin (potentially untrusted user input). Note that our taint rule allows us to skip over boilerplate policy provisioning access to shared libraries.}]
name: hello_taint
entry: /usr/bin/hello.dynamic
allow:
  # Grant read and write access to
  # /dev/tty* and /dev/pts/* devices
  - tty: rw
taint:
  # Taint after reading from
  # /dev/tty* or /dev/pts/* devices
  - tty: r
\end{listing}

\subsection{Filesystem Policy Rules}

\bpfcontain{} policy restricts access to files using a variety of rule types, each with varying degrees of granularity. The most basic, the \textit{filesystem rule}, grants access at the granularity of a given filesystem, rooted at the provided mountpoint. \textit{Subdir rules} grants recursive access to files rooted at a given directory, with a hard limit of 8 nested subdirectories. This hard limit is a technical limitation associated with the eBPF implementation, but can be adjusted at compile time. \textit{File rules} grant access at the granularity of individual files. \textit{Device rules} grant access at the granularity of commonly-used (and unprivileged) character devices, such as TTYs and the kernel's random number facilities. The policy may optionally explicitly define a specific access pattern for each filesystem object, specified using a string of access flags. For instance, read and append access to a log file would be specified as \lsi[language=yaml]|file: /var/log/mylog.log ra|.

In addition to these explicit rules, \bpfcontain{} enforces an implicit filesystem policy on all containers. For instance, \bpfcontain{} heavily restricts access to the procfs filesystem, which exposes per-process information along with certain interfaces into the kernel. A container may only access its \textit{own} per-process entries in procfs and is prohibited from accessing \textit{any} other files in procfs unless such access is explicitly marked in the container's policy file. The sysfs filesystem, which provides a similar interface into various aspects of the kernel, receives similar treatment. Overlay filesystems also receive special treatment; a container is always granted full access to an overlay filesystem belonging to its own mount namespace. In practice, this can greatly simplify the resulting security policy, since the majority of filesystem rules can simply be removed.

\subsection{Network Policy}

\bpfcontain{} confines network traffic at the socket level. \textit{Network rules} grant access to various socket operations on the IPv4 and IPv6 socket families. Socket operations are grouped by use case and partitioned into \texttt{client}, \texttt{server}, \texttt{send}, and \texttt{receive} categories accordingly. These categories may be mixed and matched according to the required level of access. An access level of \texttt{client} grants the ability to connect to a network socket, while an access level of \texttt{server} allows a container to create, bind, and shutdown a socket as well as listen for and accept connections. Similarly, \texttt{send} access grants the ability to send data (or write) to a network socket, while \texttt{receive} access grants the ability to receive data (or read) from a network socket.

Since Unix domain sockets are used for interprocess communication rather than network communication, they are handled separately by IPC policy (c.f.~\Cref{subsub:ipc_policy}). A container typically has no need for other address families beyond basic networking and inter-process communication. Therefore, in our current prototype, all other address families are implicitly denied, although this may be subject to change in future iterations.

\subsubsection{IPC Policy}\label{subsub:ipc_policy}

\bpfcontain{} enforces inter-process communication (IPC) policy at the granularity of its representation of containers. A given container can \textit{always} perform IPC between its \textit{own} processes. To enable IPC across containers, \textit{both} container security policies must specify an \textit{ipc rule} which explicitly allowlists the other policy by specifying its name. For instance, to allow IPC between \textit{policy A} and \textit{policy B}, \textit{policy A} would require an \texttt{ipc: B} rule and \textit{policy B} would require an \texttt{ipc: A} rule. At runtime, \bpfcontain{} enforces checks to make sure that both containers belong to the same IPC namespace.

\bpfcontain{} enforces its IPC policy on all inter-process communication facilities supported in the LSM infrastructure, including Unix domain sockets, signals, and SystemV IPC and shared memory.

\subsection{Capability Policy}

Over-provisioning and abuse of POSIX capabilities is endemic in extant container implementations~\cite{sultan2019_container_security,docker}. To rectify the problem of overprivilege, \bpfcontain{} strictly limits the use of POSIX capabilities by all containers, including those which have been marked default-allow. The only way to allow the use of a given POSIX capability in a \bpfcontain{} container is to explicitly allow it by adding a \textit{capability rule} with the corresponding capability.

Note that allowing the use of these capabilities is \textit{not} the same as granting additional capabilities; a process must still possess the corresponding POSIX capability in order to use it. For example, in order for a process to use the \texttt{CAP\_SYS\_ADMIN} privilege, it must both already have this privilege and be running under a policy with an explicit \texttt{CAP\_SYS\_ADMIN} capability rule. In this way, \bpfcontain{}'s capability policy can be thought of as a mask over the set of all possible capabilities.

\subsection{Implicit Policy}

\begin{lstlisting}[language=yaml,caption={Sample \bpfcontain{} policy for a container that runs Apache \texttt{httpd} together with \texttt{mysqld}. In practice, this policy could be further simplified by relying on \bpfcontain{}'s implicit overlayfs policy.},label={lst:apache}]
name: my_webapp
entry: >
  mysqld $(SQL_ARGS) &
  httpd $(HTTPD_ARGS)
allow:
  - file: /run/apache2.pid, rwd
  - subdir: /run, rc
  - subdir: /var/www, r
  - subdir: /var/www/cgi, rxm
  - subdir: /var/cache/apache2, rwc
  - subdir: /var/lib/mysql, rwc
  - subdir: /var/lib, rm
  - subdir: /var/log/apache2, rac
  - subdir: /var/log/mysql, rac
  # Network access
  - net: server, send, recv
  # Bind to privileged ports
  - capability: netBindService
  # Inter-process communication between mysqld
  # and httpd is allowed implicitly, as both
  # processes exist in the same container
\end{lstlisting}

In addition to its explicit security policy, defined in policy files, \bpfcontain{} enforces a number of \textit{implicit policies} on all containers, which are enforced regardless of configuration. In general, these policies restrict access to operations that no sane container should ever require. In total, \bpfcontain{} enforces implicit policy on 11 LSM hooks, denying access to the \texttt{bpf(2)} system call, the kernel's keyring facilities, the ptrace system call, filesystem mounts, and all of the sensitive operations gated by the kernel's lockdown LSM~\cite{lockdown}.

Besides the implicit policy enforced via LSM hooks, \bpfcontain{} takes advantage of eBPF to instrument other areas of the kernel which are not directly covered by LSM hooks. In this way, \bpfcontain{} can dynamically harden the kernel against a variety of additional container-specific attacks such as namespace escapes and host privilege escalation attacks~\cite{xin2018_container_security}. We cover the specific details of this extra enforcement in \Cref{subsub:kernel_components}.

The key advantage to these implicit policies, however, is that they make container-level policies remarkably simple.  \Cref{lst:apache} shows a sample policy for a container running both apache and mysql.  Notice that the bulk of the policy is taken up by filesystem rules, and even then the policy is relatively small, as standard activity within the container is allowed.  If we had configured an overlay filesystem (as is done by standard container runtimes) this policy could be even simpler, requiring no filesystem permissions at all.  (See~\Cref{subsection:ociintegration}.)

\section{Implementation}\label{sub:implementation}

\begin{figure}[ht]
  \centering
  \includegraphics[width=0.9\linewidth]{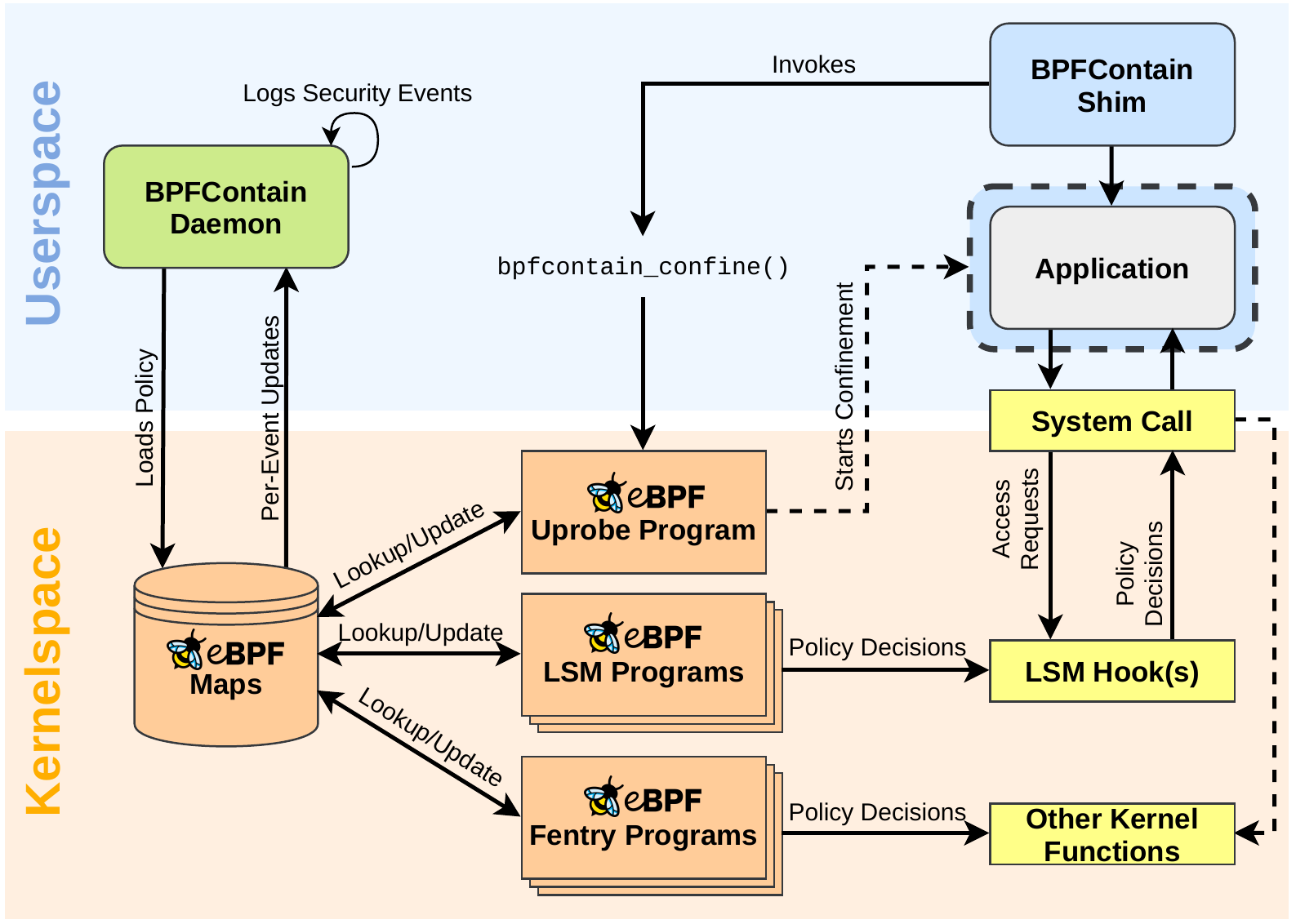}
  \caption{A detailed depiction of \bpfcontain{}'s architecture. The daemon is responsible for loading and managing \bpfcontain{}'s eBPF programs and maps, translating and loading policy into the kernel, and logging security events to userspace. eBPF programs attach to LSM hooks and other security-critical kernel functions to enforce per-container policy. A userspace shim application is responsible for launching the containerized application and starting confinement. Unlike other container implementations, the shim has no need to directly communicate with the daemon and runs with no additional privileges.}\label{fig:architecture}
\end{figure}

\bpfcontain{} consists of three functional components: (1) a privileged daemon, running in userspace; (2) a minimal shim application for launching containers; and (3) eBPF programs attached to various system events and LSM hooks. \Cref{fig:architecture} provides a detailed overview of \bpfcontain{}'s architecture. In the rest of this section, we discuss the userspace and kernelspace components respectively.

\subsection{Userspace Components}

In userspace, \bpfcontain{} consists of a privileged daemon and an unprivileged shim application used to launch containers. The daemon is responsible for loading and managing \bpfcontain{}'s eBPF maps and programs, loading per-container policy into the kernel, and logging security events for provenance. The shim's only purpose is to invoke the \texttt{bpfcontain\_confine} library call to initiate confinement followed by an \texttt{execve(2)} call to launch the correct entrypoint application.

Unlike traditional container implementations, the shim (and consequently any container launched using the shim) requires no special privileges. To obviate the need for communicating with the daemon directly, we instead instrument a uprobe (userspace probe) eBPF program on the \texttt{bpfcontain\_confine} library call. Whenever a userspace application makes this library call, it traps to the uprobe which then creates a new container for the process and associates it with the correct security policy. From the application's perspective, this procedure is totally transparent.

Both the daemon and shim are written in Rust, a fast and reliable systems programming language with strong memory- and thread-safety guarantees. \bpfcontain{} leverages libbpf-rs and BPF CO-RE (compile once, run everywhere) to load its eBPF programs and maps into the kernel. With CO-RE, \bpfcontain{} embeds its eBPF object code directly into the resulting executable, allowing for a single binary to be compiled once and executed on any system running the minimum required kernel version. We expect that this property will prove highly advantageous for cloud-based and IoT-based deployments.

\subsection{Kernelspace Components}
\label{subsub:kernel_components}

On the kernel side \bpfcontain{} leverages a variety of eBPF programs and maps to enforce policy, log security events, and track the state of running containers. In total, we instrument 38 eBPF LSM programs, two fentry (kernel function entry) programs, two scheduler tracepoints, and one uprobe (userspace probe). Taken together, these components provide a complete abstraction for containers and container behaviour in kernelspace and offer an effective mechanism for enforcing least-privilege on running containers.

\paragraph*{eBPF Maps.}

To store policy, track the state of running containers, and log security events to userspace, \bpfcontain{} employs several eBPF maps. These maps are specialized data structures which reside in the kernel and provide bi-directional (by default) lookup and update capabilities to privileged userspace applications and eBPF programs.

A processes map tracks the state of containerized processes and manages their associations with running containers. Similarly a containers map tracks the state of running containers, including policy association, information about namespace and cgroup membership, whether the container is tainted, and a reference count of how many processes are running under the container.

We additionally define one map per policy category (e.g.~file, filesystem, and network policy) and mark each policy map read-only after populating it from userspace. When a container requests access to a sensitive resource, eBPF programs query the policy maps and the state of the running container and use this information to come to a policy decision. A final specialized map acts as a ring buffer to store and forward security events (e.g.~policy denials) to userspace so that the \bpfcontain{} daemon can log them.

\paragraph*{LSM Programs.}

The majority of \bpfcontain{}'s policy enforcement mechanism is implemented using eBPF programs attached to LSM hooks. These LSM hooks are strategically positioned in various security-critical sections of kernel code and define the canonical interface for implementing custom access control policy in the kernel. These LSM programs can coexist with any other Linux security module running on the system (e.g.~AppArmor, SELinux, or Yama) and work co-operatively with other implementations to come to a final policy decision. This means that, although \bpfcontain{} is designed to completely replace existing LSMs for container security, exclusivity is not a requirement.

\paragraph*{Hot Patching Kernel Vulnerabilities.}

While LSM programs provide a strong basis for policy enforcement, we found that the kernel could benefit from additional hardening in areas which are not directly protected by LSM hooks. For instance, Xin \etal~\cite{xin2018_container_security} identified a common class of container privilege escalation attacks which work by exploiting kernel code execution vulnerabilities to force an invocation of the kernel's \texttt{commit\_creds} function. The attacker then uses this function to update their process' credentials with elevated privileges. Their original paper proposed a simple defence involving a 10 line patch to the kernel's \texttt{commit\_creds} function that adds a check to see if a namespace confines the process. If it is, assume that it is in a container, and block any updates to credentials that would result in escalation of privilege~\cite{xin2018_container_security}. While effective, this solution is inflexible in that it assumes a specific container abstraction based on namespace membership and requires an out-of-tree kernel patch.

In \bpfcontain{}, we implement a similar mitigation technique but instead use a special eBPF program typed called an fentry probe. Fentry probes replace kernel function entrypoints with a shim function that trampolines to an eBPF program. \bpfcontain{} attaches such an fentry probe to the \texttt{commit\_creds} function and uses it to check for privilege escalation within a container. If the probe detects privilege escalation, it simply kills the offending process.

We apply a similar technique to the \texttt{switch\_task\allowbreak{}\_namespaces} function to prevent a container from escaping namespace isolation. In principle, it is possible to add arbitrarily many such fentry programs to \bpfcontain{} as new kernel-level vulnerabilities are discovered. These \enquote{hot patches} can then be applied dynamically, simply by reloading the \bpfcontain{} daemon.

\paragraph*{Scheduler Tracepoints.}

To keep track of process state and container membership, we instrument tracepoint eBPF programs on the scheduler, tracking task creation and task exits. When \bpfcontain{} detects that a task has been created, it checks to see if its parent is in a \bpfcontain{} container. If this check passes, the child process is associated with the same container and we atomically increment the container's reference count. In this way, \bpfcontain{} recursively builds its own per-container process tree.

Similarly, when a task exits, we check if it belongs to a container and decrement the corresponding reference count. Once a container's reference count reaches zero, \bpfcontain{} removes it from the containers map, freeing up space for a new container to take its place.

\paragraph*{Extending the Kernel ABI with Uprobes.}

An important difference between \bpfcontain{} and traditional container implementations is that a process can containerize itself without requiring any additional privileges. We accomplish this via a shim application that makes a single library call, \texttt{bpfcontain\_confine}, before calling \texttt{execve(2)} to launch the target application. On the kernel side, \texttt{bpcontain\_confine}'s logic is implemented using a uprobe (userspace probe), which replaces the function address with a trap to an eBPF program. This eBPF program first checks to ensure that the calling process is not already associated with a container, to prevent an exiting container from escaping confinement. If this check succeeds, \bpfcontain{} creates a new container, associates it with the correct policy, adds the current process to the container, and passes control back to userspace.

\section{Evaluation}
\label{sec:eval}

Here we evaluate the security and performance of \bpfcontain{}.  We evaluate its security by analyzing its implementation in the context of the threat model presented in \Cref{sub:threat-model}, while performance is empirically analyzed relative to programs running with no confinement and as confined by AppArmor.  While neither of these evaluations are exhaustive, they show that \bpfcontain{} is strong in the context of its threat model while imposing comparable overhead to other confinement solutions.

\subsection{Security}

To prevent attacker-controlled containers from making system calls that could potentially compromise other containers or the host system, we must ensure that \bpfcontain{} acts as a reference monitor with respect to kernel operations.
According the Anderson's reference monitor model \cite{anderson1973_planning_study}, a secure reference validation mechanism must satisfy the properties of \textit{complete mediation}, \textit{tamper resistance}, and \textit{verifiability}.  \bpfcontain{}'s containment guarantees are proportional to the degree these three properties are satisfied.  Below we discuss the degree to which they hold for \bpfcontain{}.

\paragraph*{Complete Mediation.}

Recall that most of \bpfcontain{}'s policy enforcement mechanism leverages the Linux Security Modules framework exposed by the kernel. If we assume that the property of complete mediation holds for the LSM framework itself, we can say that \bpfcontain{} achieves complete mediation insofar as its LSM-level policy is concerned. Other aspects of \bpfcontain{}'s policy enforcement, such as its instrumentation of the \texttt{commit\_creds} function in the kernel, serve only to complement its LSM-level policy rather than replace it. Such extensions provide additional kernel-level hardening against attacks mounted from containers and, thus, increase the strength of \bpfcontain{}'s complete mediation guarantees. In summary, we can say that \bpfcontain{}'s complete mediation at least reduces to that of the LSM framework itself, and extends it in the best case.

\paragraph*{Tamper Resistance.}

Where feasible, \bpfcontain{} uses its own mechanisms to prevent tampering with its state while it is running.  This protection involves multiple mechanisms, described below.

\bpfcontain{} is potentially vulnerable to rogue \texttt{bpf(2)} system calls to modify or remove its code and maps.  The kernel gates access to the \texttt{bpf(2)} system call with the CAP\_SYS\_ADMIN capability, which means a process would effectively require root privileges before accessing any map on the system \cite{linux_bpf}.  To protect against compromised privileged processes, \bpfcontain{} includes specific logic in its LSM probe on the \texttt{bpf(2)} system call itself, preventing any userspace process other than the \bpfcontain{} daemon itself from directly accessing or modifying its policy maps. Additionally, maps responsible for managing process and container state are restricted using the \texttt{BPF\_F\_READONLY} flag, meaning that they can only be modified from within \bpfcontain{}'s LSM probes, and never via the \texttt{bpf(2)} system call \cite{linux_bpf}.

\bpfcontain{} should continue to enforce protections even if its userspace daemon is terminated.  Normally eBPF code is loaded as long as the associated file descriptor is open~\cite{starovoitov2018_lifetime}; thus, when file descriptors are closed on program termination, the associated eBPF code and maps are freed.  \bpfcontain{} ensures their persistence by pinning all of its maps and programs to a special filesystem called \texttt{bpffs}, thus incrementing the reference counts on the file descriptors.  \bpfcontain{} also instruments an LSM probe preventing any other process from calling \texttt{unlink(2)} on its pinned file descriptors.

\bpfcontain{} is still vulnerable to kernel-level compromises, for example through code injection attacks.  This vulnerability is no more than that of any other Linux kernel security mechanism, however.  Further, as explained above, \bpfcontain{} has significant protections against unauthorized yet privileged users and processes.  While its implementation is not quite as locked down as SELinux or AppArmor, its current implementation is hardened without a significant usability impact.

\paragraph*{Verifiability.}

While \bpfcontain{} has not been formally verified, its design and implementation do facilitate verification in multiple ways.  First, \bpfcontain{}'s eBPF code is run through the eBPF bytecode verifier whenever it is run.  One of the biggest challenges in working with eBPF is the strictness of the verifier as it imposes so many restrictions in order to ensure safety.  These very limitations, however, prevent many typical coding errors and form a solid basis for more advanced verification approaches.

\bpfcontain{} has a comparatively small kernel-level codebase (well under 2000 lines of kernelspace code) compared to conventional LSM implementations such as SELinux or AppArmor, further facilitating manual audits or automated verification.  However, given that \bpfcontain{} is ultimately only as secure as the rest of the Linux kernel, such verification can only provide modest additional security guarantees.

\subsection{Performance}
\label{subsection:performance}

To establish the performance overhead of \bpfcontain{} as a drop-in replacement for confinement solutions such as AppArmor, we evaluated its performance using several benchmarking tests provided in the Phoronix Test Suite~\cite{phoronix}, to our knowledge the most comprehensive Linux benchmarking platform. \Cref{tab:bench_tests} describes each benchmarking test in detail. To establish a baseline for comparison with AppArmor, we ran each test under five distinct configurations, described in \Cref{tab:bench_configs}. See Appendix \Cref{tab:full_results} for full results of each benchmark.

Tests were run in a KVM-enabled QEMU virtual machine running a 5.10 Linux kernel on an idle host machine. The virtual machine was given 8 virtual CPUs at 2.99GHz and 16GB of RAM\@.

\begin{table}[ht!]
  \centering
  \caption{A description of the benchmarking tests.}\label{tab:bench_tests}
  \begin{tabular}{lp{13em}}
  \toprule
  \bf Test           & \bf Description \\
  \midrule
  Apache Webserver   & Measures requests per second of serving static HTML content via the Apache \texttt{httpd} webserver. \\
  Kernel Compilation & Measures time taken to build the Linux kernel. \\
  Create Files       & Measures time taken to create, write, and delete files. \\
  Create Threads     & Measures time taken to create new threads. \\
  Create Processes   & Measures time taken to fork into new processes. \\
  Launch Programs    & Measures time taken for fork and execute a dummy program. \\
  Memory Allocations & Measures time taken allocate and free small chunks of memory (4 -- 128 bytes).\\
  \bottomrule
  \end{tabular}
\end{table}

\begin{table}[ht!]
  \centering
  \caption{A description of the system configurations during benchmarking.}\label{tab:bench_configs}
  \begin{tabular}{lp{13em}}
  \toprule
  \bf Configuration   & \bf Description \\
  \midrule
  No Security         & No LSM is running on the system. \\
  AppArmor Base       & AppArmor is running without any security profiles enabled. \\
  AppArmor Allow      & AppArmor is running and enforcing a security profile that allows all operations. \\
  \bpfcontain{} Base  & \bpfcontain{} is running without any security profiles enabled. \\
  \bpfcontain{} Allow & \bpfcontain{} is running and enforcing a security profile that allows all operations. This exercises the full code path of all BPF programs. \\
  \bottomrule
  \end{tabular}
\end{table}

\begin{figure}[ht]
  \centering
  \includegraphics[width=1\linewidth]{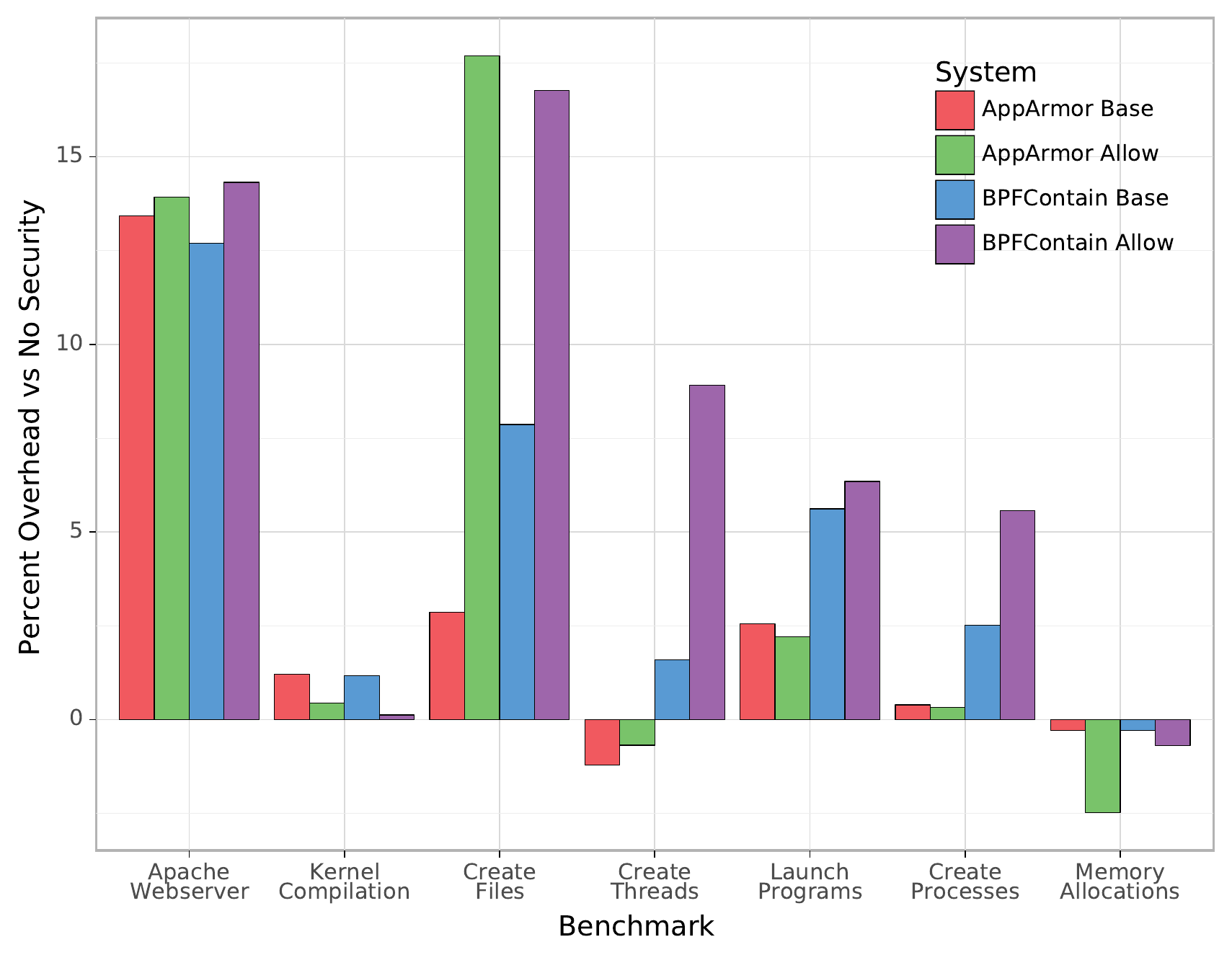}
  \caption{Percent overhead of various configurations vs no security, lower percentages are better. Negative results were found to be within margin of error.}\label{fig:benches}
\end{figure}

\Cref{fig:benches} presents the comparative percent overheads of all benchmarking tests across each system configuration, as compared to the configuration without any security enabled. Our results indicate that there is no significant difference between the overhead imposed by \bpfcontain{} and AppArmor, as all percent differences were well within the margin of error (c.f.~\Cref{tab:full_results}). Thus we conclude that \bpfcontain{} is competitive with AppArmor in terms of performance overhead on both confined and unconfined processes. Further, in all cases \bpfcontain{} exhibits less than 16\% overhead on the running system, which we find to be acceptable in practice.

\section{Related Work}
\label{sec:related}

Several researchers \cite{sultan2019_container_security,xin2018_container_security,mullinix2020_security_measures} have examined various aspects of the container security spanning various container management platforms and confinement mechanisms. Sultan \etal~\cite{sultan2019_container_security} examined the container security landscape in detail, identifying strengths, weaknesses, patterns in academic literature,  and promising future work opportunities. Their recommendations included work towards a container-specific LSM \cite{sultan2019_container_security}, which inspired the research direction for \bpfcontain{}. Xin \etal~\cite{xin2018_container_security} presented a detailed taxonomy of container security exploits and analyzed container management platforms' security against their exploit database. Mullinix \etal~\cite{mullinix2020_security_measures} presented a comprehensive analysis of Docker security and its underlying mechanisms along with the state-of-the-art solutions in academia for measuring and hardening Docker security.

Vulnerability analysis of container images \cite{shu2017_image_vuln,kwon2020_divds,brady2020_docker_cloud} has proved a lucrative technique for identifying areas of weakness in container configurations. Shu \etal~\cite{shu2017_image_vuln} presented their DIVA framework for automatic Docker Hub image vulnerability analysis and aggregated vulnerability data from over 350,000 Docker Hub images. In particular, they found that Docker images contained an average of 180 security vulnerabilities and that these vulnerabilities often propagate between parent and child images \cite{shu2017_image_vuln}. Kwon and Lee \cite{kwon2020_divds} used a similar technique in DIVDS, extracting vulnerabilities from container images and offering an interface to compare vulnerability severity and optionally add specific vulnerabilities to an allowlist. Brady \etal~\cite{brady2020_docker_cloud} applied image vulnerability scanning to a continuous integration pipeline to identify container vulnerabilities in production deployments.

Other approaches consider ways to harden container management platforms directly, using existing Linux security features. Chen \etal~\cite{chen2019_container_dos} proposed a framework for mitigating denial of service attacks against the host system mounted from containers, using a combination of cgroups and kernel-module-based enforcement to limit resource consumption. While such cgroup-based methods effectively restrict resource-based denial of service attacks, they are insufficient for implementing least-privilege. To simplify system call filtering rules and reduce overprivileged access to the host system, Lei \etal~\cite{lei2017_speaker} introduced SPEAKER to partition a container's seccomp-bpf profile into multiple execution phases. The critical insight that informed their approach was that a container's setup phase and main work loop often involve disparate sets of system calls \cite{lei2017_speaker}. Xin \etal~\cite{xin2018_container_security} proposed patching critical functions in the kernel, such as \texttt{commit\_creds} to be container-aware to mitigate the threat of kernel privilege escalation exploits mounted from containers.

Many least-privilege enforcement mechanisms for container security rely on Linux Security Modules for mandatory access control. Loukidis \etal~\cite{loukidis2018_dockersec} proposed a mechanism for automatically deriving per-container AppArmor policy based on image characteristics and runtime information gathered from individual containers. Based on the observation that different containers often have disparate security requirements, Sun \etal~\cite{sun2018_security_namespace} proposed adopting a new security namespace to allow specific containers to load their own LSM implementations, independent of the rest of the system. Citing their work as a good starting point, Sultan \etal~\cite{sultan2019_container_security} proposed that further research should be dedicated to the notion of a container-specific LSM. \bpfcontain{} represents another step towards such a container-specific LSM implementation.

\bpfcontain{} is not the first research project to propose exposing LSM hooks to userspace through eBPF. Landlock \cite{landlockio,landlock_patch} is an experimental Linux Security Module presented by Salaun to expose a subset of LSM hooks to unprivileged userspace programs. Under Landlock, userspace programs write and load eBPF programs into the kernel to filter their accesses. Unfortunately, the community has since recognized that allowing unprivileged processes to load eBPF programs into the kernel is fundamentally insecure, regardless of any limitations imposed on program type and functionality \cite{corbet2019_krsi}. Thus, Landlock has not been merged into the mainline kernel and will likely remain out-of-tree going forward. Unlike Landlock, Singh's KRSI \cite{corbet2019_krsi,singh2019_krsi} allows privileged users to attach eBPF programs to LSM hooks. Since KRSI does not require unprivileged processes to load and manage eBPF programs, it does not suffer the same fundamental security issues that have detracted from Landlock.

While KRSI serves as the infrastructure for implementing LSM programs in eBPF, developers must still provide their own implementations for any eBPF LSM hooks they wish to use. Findlay \etal~\cite{findlay2020_bpfbox} introduced bpfbox as the first full process confinement mechanism using these eBPF LSM hooks.  \bpfcontain{} differs from bpfbox in three key ways: 1) \bpfcontain{} confines containers (sets of processes and associated resources) while bpfbox confines individual processes; this also affords a significant simplification of security policy, since implicit rules can be enforced at the per-container granularity 2) \bpfcontain{} is implemented using BPF CO-RE and Rust rather than bcc and Python, greatly reducing it storage and runtime overhead, and 3) \bpfcontain{} moves beyond just LSM-layer enforcement to apply additional kernel hardening against privilege escalation attacks that can undermine its protection.

\section{Discussion}\label{sec:discussion}

When designing security solutions, we are often at the mercy of past design decisions.  One of the core insights of computer security is that it is easier to build security in rather than add it afterwards.  Central to this idea is that many security problems arise at the architectural level, and it is hard to change the architectures of deployed systems.  As a result, we often find ourselves integrating security mechanisms in fundamentally non-optimal contexts, simply because that is where they can be implemented.

\bpfcontain{} is a demonstration of how operating system extensibility can enable the development of new security abstractions that more closely match the problems at hand.  Domain-specific policy language no longer has to be implemented using existing (general-purpose) security mechanisms; instead, we can implement security policies using the fundamental functional abstractions of the operating system.

\subsection{Future Directions}
\paragraph*{Integration with Existing Container Runtimes.}\label{subsection:ociintegration}
Our focus with \bpfcontain{} has been on isolation rather than virtualization. A potential next step would be to integrate \bpfcontain{} with Docker~\cite{docker}, Kubernetes~\cite{kubernetes}, or OpenShift~\cite{openshift}.  Given that these container runtimes are largely Open Container Initiative (OCI) compliant~\cite{oci}, it should be possible to integrate with them in a portable fashion. Further, since eBPF can transparently instrument userspace code, such integration would in principle be possible without making changes to the source code of the underlying container runtime.  Policy generation and enforcement could be tied directly with container configuration, offering streamlined, precise, yet simple confinement specifications that fail closed rather than fail open.

\paragraph*{Rootless Containers in eBPF.}\label{subsection:rootless}
Another potential path would be to add virtualization and management capabilities directly to \bpfcontain{}.  The advantage of doing so would potentially be greatly simplified userspace tools due to enhanced kernel-level functionality implemented through new eBPF helpers.  These helpers could, for instance, be used to transparently move process groups into new namespaces and cgroups or manage filesystem mounts within a mount namespace, transparently to the target application. \bpfcontain{} could integrate these helpers into its container lifecycle management probes to enforce namespace, cgroup, and mount policies as well. Not only would such an extension enable fully application-transparent namespace and cgroup management, but it would also obviate the need for the root privileges currently required by container management systems.

On the policy language side, the integration of virtualization with \bpfcontain{}'s enforcement mechanisms presents opportunities for streamlining the configuration and policy associated with \bpfcontain{} containers. For instance, filesystem and mount namespace rules could be combined into one explicit \texttt{mount} rule. Under a given mount rule, \bpfcontain{} would mount an overlay filesystem in the container's mount namespace and automatically allow access to this mounted filesystem in its LSM policy. This new integration would not only significantly streamline container configuration, but it would also reduce complexity in filesystem and file rules. For example, one mount rule could replace a series of file rules specifying access to required shared libraries.  Thus, by further refining the abstractions presented to userspace by the kernel, we can enable solutions that are simpler, easier to configure, and potentially much more secure.

\paragraph*{Running \bpfcontain{} without the Daemon}After the \bpfcontain{} daemon loads its eBPF programs and maps into the kernel, its only remaining purposes is to log security events to userspace. With the ability to pin eBPF objects (thus maintaining a reference count), it would in principle be possible to completely remove the need for the daemon altogether. Naturally, this would significantly reduce \bpfcontain{}'s attack surface, and this would mark an obvious improvement over existing container runtimes like Docker and Kubernetes which rely on privileged daemons in order function correctly. While, in the current implementation, disabling the \bpfcontain{} daemon would disable its audit logging capabilities, it may be possible to make use of a new eBPF iterator program type added in recent kernels to implement similar functionality.

\subsection{Prototype Limitations}
While we believe \bpfcontain{} is a promising approach to container confinement, in its current form it has some limitations.  One is that there are hard limits on policy complexity due to the current limits of eBPF\@. eBPF's maps must be constrained to some fixed size that is determined at map creation time. Since \bpfcontain{} uses eBPF maps to store per-container policy, the maximum map size bounds the number of possible rules for each policy category. This upper bound is not strictly an issue since \bpfcontain{} loads policy when the daemon first starts, and thus map sizes can be predetermined based on policy files' contents. However, it would be desirable to add dynamically loadable policy into future versions, which would require handling this map size restriction at runtime.  Given eBPF's strong safety guarantees are due in part to draconian restrictions on dynamic memory allocation, removing this limitation is non-trivial.  Work on garbage collected map types, however, may solve this issue in the near future.

\section{Conclusion}
\label{sec:conclusion}

This paper presented \bpfcontain{}, a novel least-privilege implementation for container security. \bpfcontain{} exposes a simple YAML-based policy configuration language to userspace that conforms to existing container management mechanisms' semantics while supporting ad-hoc confinement use cases through high-level policy rules and optional default-allow enforcement. Because \bpfcontain{} is written in eBPF, it can be deployed on virtually any system running a recent Linux kernel with no kernel-level changes and none of the risks of out-of-tree kernel modules.  eBPF also allows \bpfcontain{} to protect itself against privilege escalation exploits that could prevent its functioning.

When integrated with OCI-compliant container management systems, \bpfcontain{} will provide strong yet flexible container confinement, enabling more secure multi-tenant container deployments.  It also helps demonstrate the potential of using eBPF to extend the Linux kernel with new security abstractions.

\printbibliography

\appendix

\section{Appendix A: Benchmarking Results}
\Cref{tab:full_results} lists detailed benchmark results comparing \bpfcontain{} to AppArmor.

\begin{table*}[h!]
\centering
\caption{Full benchmarking results comparing AppArmor and \bpfcontain{}.}\label{tab:full_results}
\begin{tabular}{lllrrl}
\toprule
              Test & System           & Units   & Mean     & Stdev  & Percent Overhead \\
\midrule
  Apache Webserver & Base             & req/sec & 23258.19 & 108.03 & --- \\
  Apache Webserver & BPFContain Base  & req/sec & 20306.59 & 326.57 & 12.69\% \\
  Apache Webserver & BPFContain Allow & req/sec & 19928.84 & 203.66 & 14.31\% \\
  Apache Webserver & AppArmor Base    & req/sec & 20134.33 & 534.40 & 13.43\% \\
  Apache Webserver & AppArmor Allow   & req/sec & 20020.81 & 244.99 & 13.92\% \\
Kernel Compilation & Base             & sec     & 188.88   & 1.73   & --- \\
Kernel Compilation & BPFContain Base  & sec     & 191.10   & 1.48   & 1.17\% \\
Kernel Compilation & BPFContain Allow & sec     & 189.12   & 2.08   & 0.13\% \\
Kernel Compilation & AppArmor Base    & sec     & 191.18   & 1.59   & 1.21\% \\
Kernel Compilation & AppArmor Allow   & sec     & 189.72   & 1.53   & 0.44\% \\
      Create Files & Base             & usec    & 19.32    & 0.14   & --- \\
      Create Files & AppArmor Base    & usec    & 19.87    & 0.18   &  2.86\% \\
      Create Files & AppArmor Allow   & usec    & 22.74    & 0.21   & 17.69\% \\
      Create Files & BPFContain Allow & usec    & 22.56    & 0.15   & 16.76\% \\
      Create Files & BPFContain Base  & usec    & 20.84    & 0.22   &  7.87\% \\
    Create Threads & Base             & usec    & 21.48    & 1.74   & --- \\
    Create Threads & AppArmor Base    & usec    & 21.22    & 1.31   & -1.21\% \\
    Create Threads & AppArmor Allow   & usec    & 21.33    & 1.30   & -0.68\% \\
    Create Threads & BPFContain Allow & usec    & 23.39    & 1.20   &  8.91\% \\
    Create Threads & BPFContain Base  & usec    & 21.82    & 1.48   &  1.59\% \\
   Launch Programs & Base             & usec    & 61.35    & 0.55   & --- \\
   Launch Programs & AppArmor Base    & usec    & 62.92    & 0.66   & 2.55\% \\
   Launch Programs & AppArmor Allow   & usec    & 62.71    & 0.71   & 2.21\% \\
   Launch Programs & BPFContain Allow & usec    & 65.24    & 0.58   & 6.35\% \\
   Launch Programs & BPFContain Base  & usec    & 64.80    & 0.47   & 5.62\% \\
  Create Processes & Base             & usec    & 41.33    & 2.28   & --- \\
  Create Processes & AppArmor Base    & usec    & 41.49    & 1.28   & 0.40\% \\
  Create Processes & AppArmor Allow   & usec    & 41.47    & 1.89   & 0.33\% \\
  Create Processes & BPFContain Allow & usec    & 43.63    & 1.82   & 5.57\% \\
  Create Processes & BPFContain Base  & usec    & 42.37    & 2.29   & 2.52\% \\
Memory Allocations & Base             & nsec    & 93.80    & 0.40   & --- \\
Memory Allocations & AppArmor Base    & nsec    & 93.53    & 0.39   & -0.30\% \\
Memory Allocations & AppArmor Allow   & nsec    & 91.48    & 0.28   & -2.48\% \\
Memory Allocations & BPFContain Allow & nsec    & 93.15    & 0.62   & -0.70\% \\
Memory Allocations & BPFContain Base  & nsec    & 93.54    & 0.29   & -0.29\% \\
\bottomrule
\end{tabular}
\end{table*}

\end{document}